\documentclass[3p,times,twocolumn]{elsarticle}
 \biboptions{comma,sort&compress}
 
\usepackage{graphicx}
\usepackage{here}
\usepackage{ecrc}

\volume{00}

\firstpage{1}

\journalname{Nuclear and Particle Physics Proceedings}

\runauth{}

\jid{nppp}

\jnltitlelogo{Nuclear and Particle Physics Proceedings}

\usepackage{amssymb}

\usepackage[figuresright]{rotating}

\begin{document}

\begin{frontmatter}

\title{$1^+$ XTZ States Within QCD Sum Rules}
 \cortext[cor0]{Talk given at 25th High-Energy Physics 
 International Conference in Quantum Chromodynamics (QCD 22), 
 4-7 July 2022, Montpellier-FR}
\author[label1]{R.M. Albuquerque\fnref{fn1}}
\fntext[fn1]{Speaker.}
\ead{raphael.albuquerque@uerj.br}
\address[label1]{Faculty of Technology, Rio de Janeiro 
State University (FAT,UERJ), Brazil.}
\author[label2,label3]{S. Narison}
\ead{snarison@yahoo.fr}
\address[label2]{Laboratoire Univers et Particules 
de Montpellier, CNRS-IN2P3, Case 070, Place Eug\`ene
Bataillon, 34095 - Montpellier, France.}
\address[label3]{Institute of High-Energy Physics of 
Madagascar (iHEPMAD), University of Ankatso, 
Antananarivo 101, Madagascar.}
\author[label3]{D. Rabetiarivony}
\ead{rd.bidds@gmail.com}

\pagestyle{myheadings}
\markright{ }
\begin{abstract}
{ We present improved estimates of the couplings, masses 
and mass ratios of the $X_Q, Z_Q$ and $T_{QQ\bar q\bar q'}$ 
states ($Q\equiv c,b~;~q,q'\equiv u,d,s$) using (inverse) 
QCD Laplace sum rules (LSR), their ratios ${\cal R}$ and 
double ratios (DRSR), within stability criteria. We conclude 
that the observed $X_c(3872)$ and $Z_c(3900)$ are 
{\it tetramoles} states (superposition of quasi-degenerated 
molecule and tetraquark states having similar couplings to 
the currents) with the predicted masses\,:
$M_{{\cal T}_{X_c}}=3876(44)$ MeV and 
$M_{{\cal T}_{Z_c}}=3900(42)$ MeV. We also do an extensive 
analysis of the four-quark nature of different 
$T_{QQ\bar q\bar q'}$ axial-vector states. Then, combining 
${\cal R}$ and DRSR, we reanalyze the observed state $X_c(3872)$ 
and we obtain a precise prediction of $M_{T_{cc}^{1^+}}$=3886(6) MeV. 
Extending to the beauty sector, we find the results: 
$M_{{\cal T}_{Z_b}}=10579(99)$ MeV and $M_{X_b}=10545(131)$ MeV. 
Finally, we confront our combined LSR $\oplus$ DRSR results with the 
ones from some other approaches (lattices and quark models).}
\end{abstract}
 

\end{frontmatter}

\section{Introduction}
In this work, we estimate the couplings, masses, and mass ratios of the 
$X_Q, Z_Q$ and $T_{QQ\bar q\bar q'}$ states, using QCD spectral sum rule 
techniques (QSSR), see details in Ref.\cite{NPA2022}. Motivated by 
the recent LHCb discovery of a $1^{+}$ state at 3875 MeV\,\cite{LHCb4}, we
improve the existing QSSR results by combining the direct mass determinations
from the ratios ${\cal R}$ of Inverse Laplace sum rule (LSR) with the 
ratio of masses from the double ratio of sum rules (DRSR). The $1^+$ state 
lies just below the $D^*D$ threshold, which is a good isoscalar ($I=0$) 
$T_{cc\bar u\bar d}$ axial-vector $(J^P=1^+)$ candidate. We start our 
analysis by improving the previous estimate of mass and coupling of the 
$X_c(3872)$ which will serve as an input in our DRSR approach. We evaluate 
the SU3 breakings for $T_{cc\bar s\bar s}$ and $T_{cc\bar s\bar u}$ states. 
Finally, we extend the whole study to the b-sector including the 
$T_{bb\bar q\bar q'}$ states. Our results are confronted with the existing 
LSR results and the ones from some other approaches.

{\small
\begin{table*}[hbt]
\setlength{\tabcolsep}{0.3pc}
    {\small
\begin{tabular*}{\textwidth}{@{}c@{\extracolsep{\fill}}cl}
&\\
\hline
\hline
States &$I(J^P)$ & $\bar 3_c3_c$ Four-quark Currents\\
\hline
$Z_c$ &$(1^{+})$
&$ {\cal O}_{A_{cq}}=\epsilon_{ijk}\epsilon_{mnk}\big{
[}(q^T_i\,C\gamma_5\,c_j)(\bar q'_m \gamma_\mu C\, \bar c_n^T) + 
b \,(q^T_i\,C\,c_j)(\bar q'_m\gamma_\mu \gamma_5 C\, \bar c_n^T)\big{]}$\\
&&$ {\cal O}_{D^*_qD_q}=(\bar c\gamma_\mu q)(\bar q'\,i\gamma_5c)$  \\ \\
$X_c $&$(1^{+})$
&$ {\cal O}^3_{X} = \epsilon_{i j k} \:\epsilon_{m n k} \big{[}\left(
  q_i^T\, C \gamma_5 \,c_j \right) \left( \bar{c}_m\, \gamma^\mu
  C \,\bar{q}_n^T\right) +  \left( q_i^T\, C \gamma^\mu \,c_j \right) 
  \left( \bar{c}_m\, \gamma_5 C \,\bar{q}_n^T\right)\big{]}$\\
 && $ {\cal O}^6_{X} = \epsilon_{i j k} \:\epsilon_{m n k}\big{[} \left(
  q_i^T\, C \gamma_5 \lambda_{ij}^a\,c_j \right) \left( \bar{c}_m\, \gamma^\mu
  C\lambda_{mn}^a \,\bar{q}_n^T\right) + \left(
  q_i^T\, C \gamma^\mu \lambda_{ij}^a\,c_j \right) \left( \bar{c}_m\, \gamma_5
  C \lambda_{mn}^a\,\bar{q}_n^T\right)\big{]}$ \\
  && $ {\cal O}_{D^*_qD_q}= \frac{1}{\sqrt{2}}\big{[}(\bar q\gamma_5 c) 
  (\bar c\gamma_\mu q) - (\bar q\gamma_\mu c) (\bar c\gamma_5 q)\big{]} $ \\
  && $ {\cal O}_{\psi\pi} = (\bar c\gamma_\mu\lambda^a c)
  (\bar q\gamma_5\lambda^a q)$\\ \\
 $T_{cc\bar u\bar d}$&$0(1^{+})$
 &  $ {\cal O}_T^{1^+} = \frac{1}{\sqrt{2}}\epsilon_{i j k} 
 \:\epsilon_{m n k} \left(c_i^T\, C \gamma^\mu \,c_j \right) \big{[} 
 \left( \bar{u}_m\, \gamma_5 C \,\bar{d}_n^T\right) - \left( 
 \bar{d}_m\, \gamma_5 C \,\bar{u}_n^T\right)\big{]}$\\
$T_{cc\bar u\bar s}$&$\frac{1}{2}(1^{+})$
 &  $ {\cal O}_{T^{1^+}_{us}} = \epsilon_{i j k} \:\epsilon_{m n k}
    \left( c_i \, C \gamma^{\mu } c_j^T \right) 
    \left( \bar{u}_m \,\gamma_5 C \bar{s}_n^T \right)$\\
$T_{cc\bar u\bar d}$&$1(0^{+})$
 &  $ {\cal O}_T^{0^+} = \frac{1}{\sqrt{2}}\epsilon_{i j k} \:
 \epsilon_{m n k} \left(c_i^T\, C \gamma^\mu \,c_j \right) \big{[} 
 \left( \bar{u}_m\, \gamma_\mu C \,\bar{d}_n^T\right) + \left( 
 \bar{d}_m\, \gamma_\mu C \,\bar{u}_n^T\right)\big{]}$\\
$T_{cc\bar u\bar s}$&$\frac{1}{2}(0^{+})$
 &  $ {\cal O}_{T^{0^+}_{us}} = \epsilon_{i j k} \:\epsilon_{m n k}
    \left( c_i \, C \gamma_{\mu } c_j^T \right) 
    \left( \bar{u}_m \,\gamma^{\mu } C \bar{s}_n^T \right)$\\
$T_{cc\bar s\bar s}$&$0(0^{+})$
 &  $ {\cal O}_T^{0^+} = \epsilon_{i j k} \:\epsilon_{m n k}
    \left( c_i \, C \gamma_{\mu } c_j^T \right) 
    \left( \bar{s}_m \,\gamma^{\mu } C \bar{s}_n^T \right)$\\
   \hline\hline
  \vspace*{-0.5cm}
\end{tabular*}}
 \caption{Interpolating operators describing the $Z_c,X_c,
 T_{cc\bar q\,'\bar q}$ states discussed in this paper where $b=0$ is the 
 optimized mixing parameter\,\cite{NPA2022}.}
\label{tab:current}
\end{table*}
}

\section{The QCD Inverse Laplace Sum Rules (LSR)}
We shall work with the Finite Energy version of the 
QCD Inverse Laplace sum rules (LSR) and their ratios
\cite{SVZa,SVZb,SNB1,SNB2,SNB3,SNB4,IOFFEb,RRY,DERAF,
BERTa,YNDB,PASC,DOSCH}:
\begin{eqnarray}
    \hspace*{-0.2cm} {\cal L}^Q_n(\tau,\mu) &=&
    \int\limits_{t_Q}^{t_c} dt ~t^n~
    e^{-t\tau} ~\frac{1}{\pi} ~\mbox{Im}~\Pi^Q_1(t,\mu)~,
    \\ \nonumber \\
    {\cal R}^Q(\tau) &=& 
    \frac{{\cal L}^Q_{n+1}} {{\cal L}^Q_{n}},
\label{eq:lsr}
\end{eqnarray}
where $\tau$ is the LSR variable, $n$ is the degree of moments, 
$t_Q = \sqrt{2M_Q+m_q+m_{q'}}$ and $t_c$ is the threshold of the 
``QCD continuum". 

The spectral function $\Pi^Q_{1}(t,\mu)$ is evaluated by the calculation 
of the axial-vector correlator defined as:
\begin{eqnarray}
    \hspace*{-0.1cm} \Pi_{\mu\nu}(q^2)\hspace*{-0.2cm} &=& \hspace*{-0.2cm}
    i \hspace*{-0.05cm}
    \int \hspace*{-0.15cm}d^4x ~e^{-iqx}\langle 0\vert T [
    {\cal O}^J_{\mu}(x) ~ {\cal O}^{J ~\dagger}_{\nu}(0) ] 
    \vert 0\rangle~ \nonumber\\
    &\equiv& \hspace*{-0.2cm} -\left( g_{\mu\nu}-
    \frac{q_\mu q_\nu}{q^2}\right)
    \Pi_{1}(q^2)+\frac{q_\mu q_\nu}{q^2} \Pi_{0}(q^2)
 \label{eq:2point}
\label{eq:2-pseudo}
\end{eqnarray}
where ${\cal O}^{J}_{\mu}(x)$ are the local hadronic operators 
(see Table\,\ref{tab:current}). One can deduce the hadronic mass 
from the ratio of LSR at the optimization point $\tau_0$\,:
\begin{equation}
    {\cal R}^Q_{\cal H}(\tau_0)= M_{\cal H}^2.
\end{equation}
We shall also work with the double ratio of sum rule (DRSR)\,\cite{DRSR88}:
\begin{equation}
    r_{{\cal H'}/{\cal H}}(\tau_0) ~\equiv~ \sqrt{\frac{
    {\cal R}^Q_{\cal H'}}{{\cal R}^Q_{\cal H}}} ~=~ 
    \frac{M_{\cal H'}}{M_{\cal H}},
\end{equation}
which can be free from systematic errors provided that ${\cal R}^c_{\cal H}$ 
and ${\cal R}^c_{\cal H'}$ optimize at the same values of $\tau$ and 
of $t_c$. In this paper, we extend the previous analysis to improve 
the existing mass predictions of the $X_Q, Z_Q$ and $T_{QQ\bar q\bar q}$ 
states and give a correlation among them. We also predict the 
mass-splittings due to SU3 breakings and to spin and parity for the 
$T_{QQ\bar q\bar q'}$ states.

\section{The stability criteria for extracting the optimal 
results in LSR\label{sec:stability}}
In the LSR analysis, we have three external variables: the LSR variable 
$\tau$, the QCD continuum threshold $t_c$, and the subtraction 
point $\mu$. One considers that physical observables, like the masses and 
meson couplings, should be independent/minimal sensitive on these 
parameters. 
\begin{itemize}
    \item {\bf the $\tau$-stability:} we expect to establish a region 
    around a minimum or inflection point corresponding to the dominance of 
    the lowest ground-state contribution and the OPE convergence.
    \item {\bf $t_c$-stability:} this is a free parameter in the LSR analysis
    though one expects it to be around the mass of the first excitation. 
    To be conservative, we consider the optimal values for $t_c$ from the 
    beginning of $\tau$-stability until the LSR results start to be 
    independent of $t_c$ choice.
    \item {\bf $\mu$-stability:} used to fix, in a rigorous optimal way, 
    the arbitrary subtraction constant appearing in the perturbative 
    calculation and in the QCD input renormalized parameters. 
\end{itemize}

\subsection{The Interpolating Operators}
We shall be concerned with the lowest dimension interpolating currents 
of the $1^+$ four-quark states given in Table\,\ref{tab:current}. 
The lowest order (LO) perturbative (PT) QCD expressions, including the 
quark and gluon condensates contributions up to dimension-six condensates 
of the corresponding two-point spectral functions, the NLO PT corrections, 
the QCD input parameters and further details of the QSSR calculations are 
given in Ref.\cite{NPA2022}.

\section{The $T_{cc\bar u\bar d}\equiv T_{cc}$  $(1^{+})$ state}
Since, the pioneering work of\,\cite{LEE}, the mass and coupling 
of $T_{cc\bar q\bar q'}$ and its beauty analogue have been extracted 
from LSR by different groups\,\cite{DRSR11,WANG-Ta,WANG-Tb,ZHU-T,
AGAEV-T,MALT-T}. In this paper, we improve and extend the analysis 
in\,\cite{DRSR11} using LSR and DRSR by including the factorized NLO PT 
contributions and by paying more carefully attention on the different 
sources of the errors.  In the following, we shall consider the 
four-quark currents given in Table\,\ref{tab:current}.

\subsection*{Mass and decay constant from LSR at NLO}
\begin{figure}[hbt]
\begin{center}
\includegraphics[width=6cm]{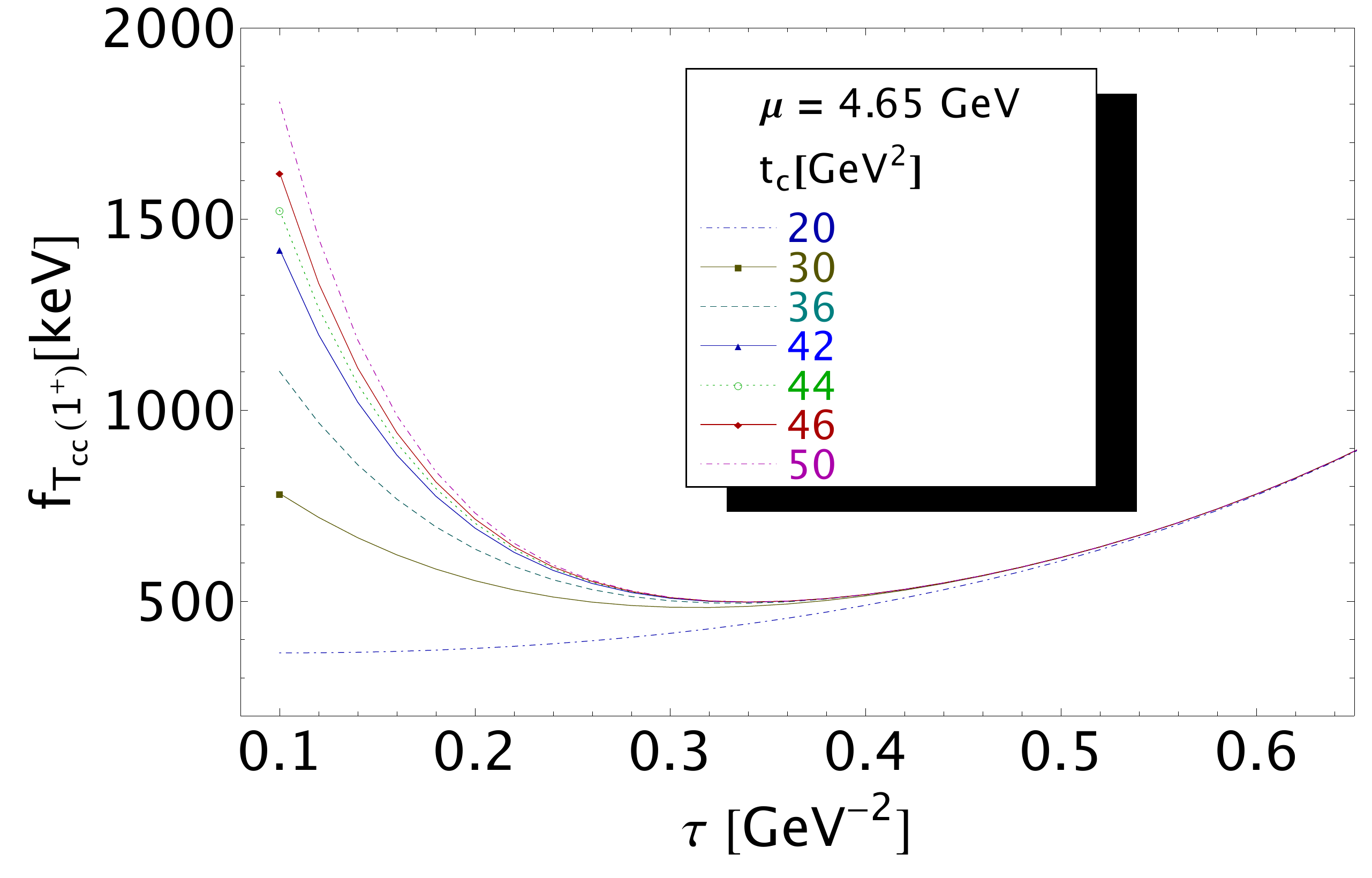}
\hspace*{0.65cm}\includegraphics[width=6.4cm]{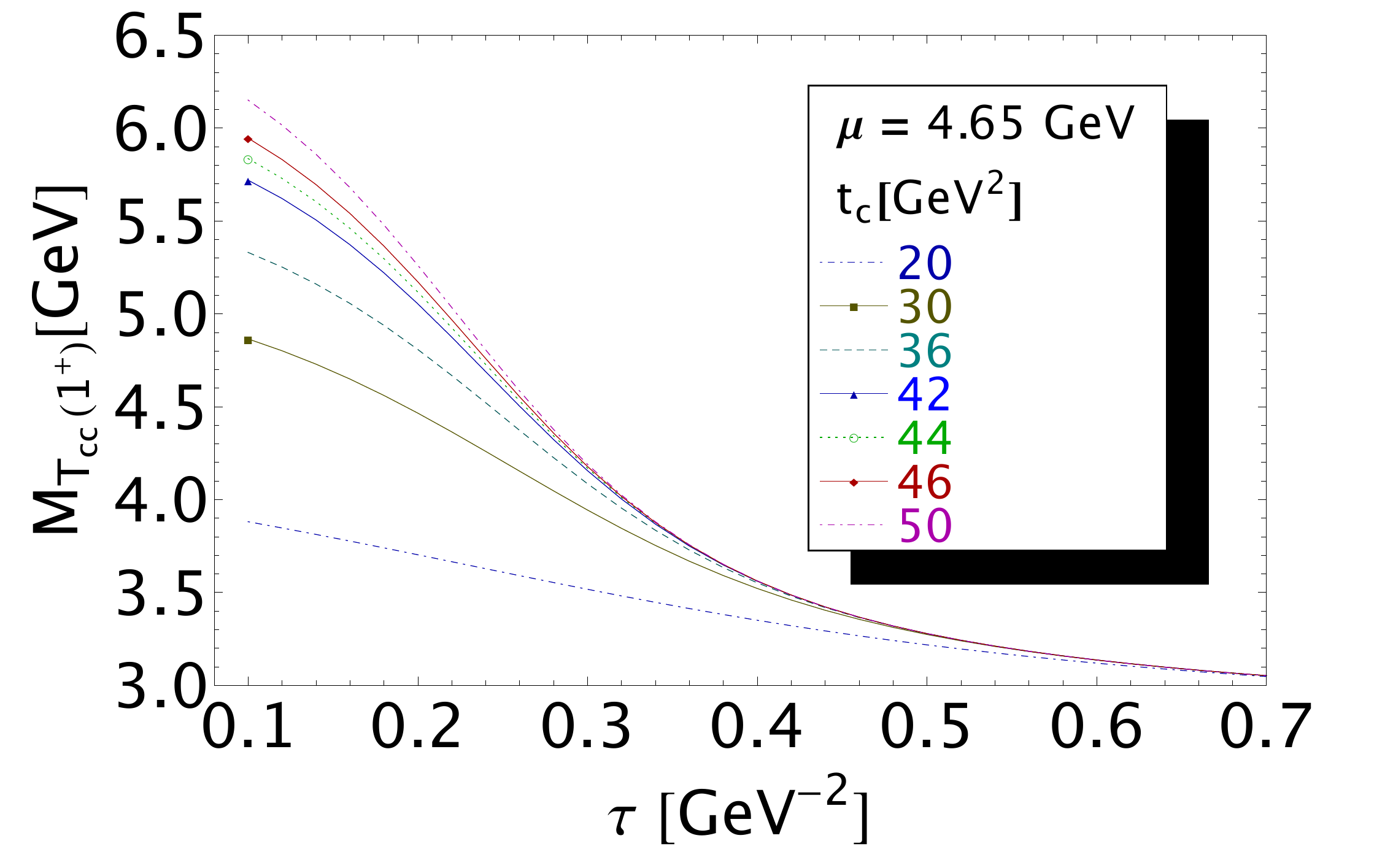}
\vspace*{-0.2cm}
\caption{\footnotesize  $f_{T^{1^+}_{cc}}$ and $M_{T^{1^+}_{cc}}$ 
as function of $\tau$ for different values of $t_c$, $\mu=4.65$ GeV 
and the usual QCD inputs.} 
\label{fig:tpcc}
\end{center}
\vspace*{-0.5cm}
\end{figure}

The $\tau$ and $t_c$ behaviours are shown in Fig.\,\ref{fig:tpcc}. The 
stability region (minimum in $\tau$ for the coupling and inflexion point 
for the mass) is obtained for the sets $(\tau,t_c)$=(0.31,30) to (0.34,46) 
in units of (GeV$^{-2}$, GeV$^2$) from which we deduce:
\begin{eqnarray}
f_{T_{cc}}(1^{+})&=&491(48)~{\rm KeV}, \nonumber\\
M_{T_{cc}}(1^{+})&=&3885(123)~{\rm MeV}~,
\label{eq:mt+cc}
\end{eqnarray}
where the mass can be compared with the experimental value 
$M_{T_{cc}}(1^{+})=3875$ MeV\,\cite{LHCb4}.

\subsection*{Ratio of masses $r_{T^{1^+}_{cc}/X_c}$ from DRSR}
The optimal result is obtained for the sets 
$(\tau,t_c)$=(1.24,15) to (1.30,20) in units of (GeV$^{-2}$, GeV$^2$) and 
we obtain the results:
\begin{eqnarray}
    r_{T^{1^+}_{cc}/X_c}&=&1.0035(10) \nonumber \\
    M_{T_{cc}}(1^{+})&=&3886(4)~{\rm MeV}
    \label{eq:dr:mt+cc}
\end{eqnarray}
where we have taken the experimental mass of the $X_c(3872)$\,\cite{PDG}. 
The result is in perfect agreement with the direct mass determination in 
Eq.\,(\ref{eq:mt+cc}) but very accurate as the DRSR is less affected by 
systematic errors which tend to cancel out. 

\subsection*{Final prediction for $M_{T_{cc}}(1^{+})$}
Using the mean of the previous results obtained in Eqs. (\ref{eq:mt+cc}) 
and (\ref{eq:dr:mt+cc}), we get the mass:
\begin{eqnarray}
    M_{T_{cc}}(1^{+})= 3886(4)~{\rm MeV}.
\label{eq:1+cc}
\end{eqnarray}
This value is comparable with the recent LHCb data $T_{cc}(1^{+})=3875$ MeV 
which is $(9\pm 4)$ MeV above the $D^*D$ threshold of 3877 MeV\,\cite{PDG}. 

\section{The $T_{cc\bar s\bar u}(1^{+})$   mass}
We extend our analysis to the $T^{1^+}_{cc\bar s\bar u}$ state, respecting 
our stability criteria and evaluating the SU3 ratio of masses 
$T^{1^+}_{cc\bar s\bar u}/T^{1^+}_{cc}$. The optimal result is given 
by the mean from the LSR at NLO and DRSR calculations: 
\begin{equation}
    M_{T_{cc\bar s\bar u}}(1^{+})=3931(7)~{\rm MeV}~.
\end{equation}

\section{The $Z_c (1^{+})$ state}
The extraction of the $Z_c$ mass has been discussed in details in 
Ref.\,\cite{Zc} using the current in Table\,\ref{tab:current}. The 
results for a $D^*D$ molecule and a four-quark state configurations 
are \cite{Zc,MOLE16}:
\begin{eqnarray}
    M_{D^*D} &=& 3912(61)~{\rm MeV} \nonumber\\
    M_{A_{cd}} &=& 3889(58)~{\rm MeV},
    \label{eq:zc}
\end{eqnarray}
which are values almost degenerated. Then, we use the DRSR for studying 
the ratio masses of these two configurations, and the optimal result 
is given by: 
\begin{eqnarray}
    r_{A_{cd}/D^*D} &=& 0.9981(6) \nonumber \\
    M_{A_{cd}} &=& 3905(61)~{\rm MeV},
\end{eqnarray}
which consolidates the previous result in Eq. (\ref{eq:zc}) from a 
direct determination. Noting in\,\cite{Zc,Zb} that the molecule $D^*D$ 
and the four-quark states are almost degenerated and have almost the same 
coupling to their respective current, we expect the physically observed 
state to be their mean which we named {\it tetramole} (${\cal T}_{Z_c}$). 
One obtains\,:
\begin{eqnarray}
    M_{{\cal T}_{Z_c}} &=& 3900(42)~{\rm MeV} \nonumber\\
    f_{{\cal T}_{Z_c}} &=& 155(11)~{\rm keV}.
\end{eqnarray}
which coincides with the experimental $Z_c(3900)$ mass.

\section{The $X_c (1^{+})$ state}
\subsection*{Mass and decay constant from the ${\cal O}^3_X$ 
current using LSR}
The mass and coupling of the $X_c (1^{+})$ have been extracted to 
lowest order (LO) \,\cite{DRSR11,DRSR07,DRSR11a} using the interpolating 
four-quark currents given in Table\,\ref{tab:current}, a molecule $D^*D$ 
and a molecule $J/\psi\pi$. These early results have been improved 
in\,\cite{MOLE16} for the ${\cal O}^3_T$ current by including NLO PT 
corrections in order to justify the use of the running heavy quark mass 
of the $\overline{MS}$-scheme in the analysis. Using the same optimization 
procedure for extracting couplings and masses, we obtain :
\begin{eqnarray}
    M_{X_{c,3}} &=& 3876(76) ~\rm MeV  \nonumber\\
    f_{X_{c,3}} &=& 183(16)~{\rm keV},
    \label{eq:x3}
\end{eqnarray}
where $f_{X_{c,3}}$ is normalized as $f_\pi=131$ MeV. One can notice the 
remarkable agreement of the central value of the mass with the data 
3871.69(17) MeV\,\cite{PDG}.

\subsection*{Mass from the ${\cal O}^6_X$ current using DRSR}
\begin{figure}[hbt]
\begin{center}
\includegraphics[width=6cm]{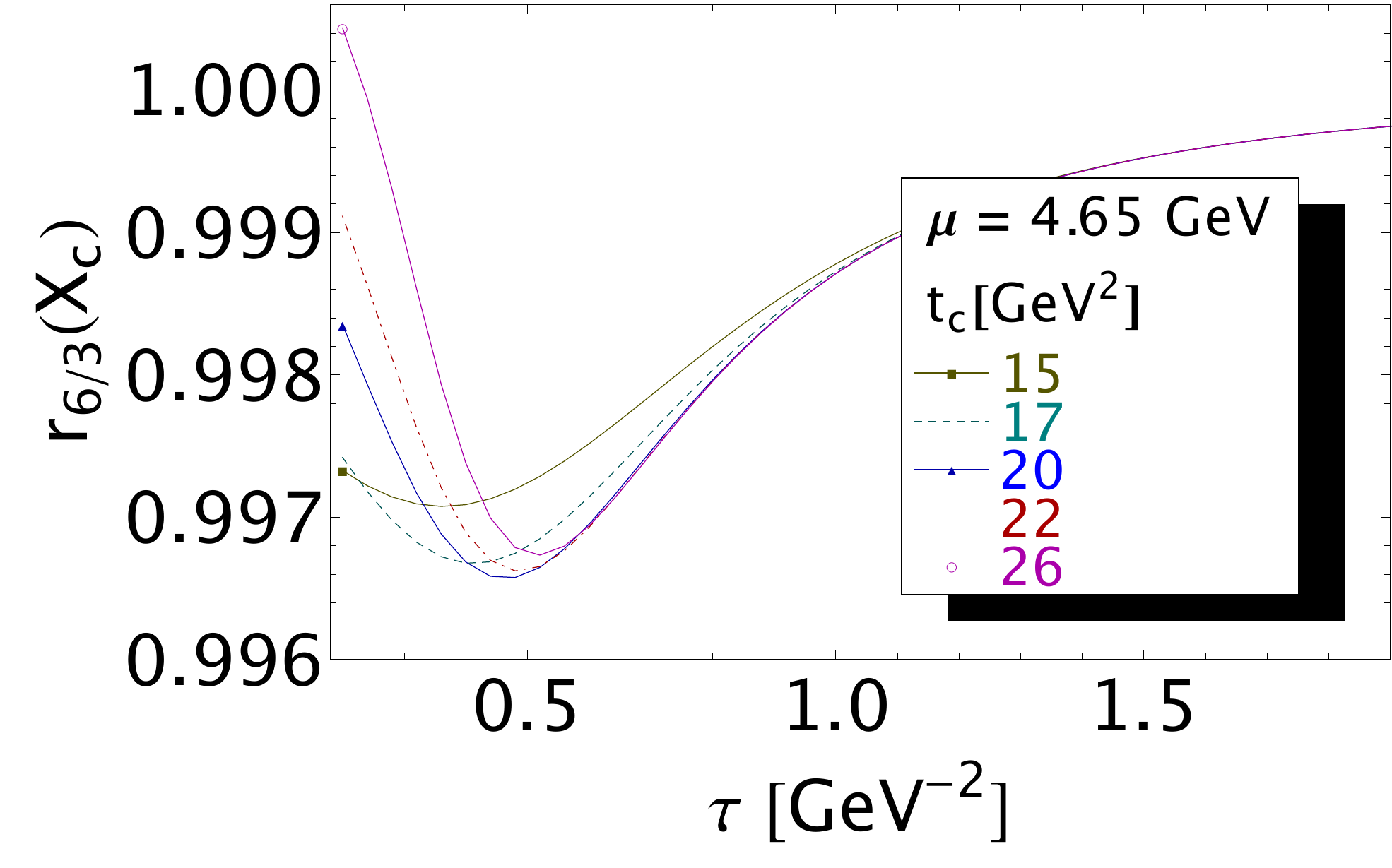}
\vspace*{-0.2cm}
\caption{\footnotesize  $r_{6/3}$  as function of $\tau$ at NLO for 
different values of $t_c$, for $\mu$=4.65 GeV and for the usual QCD 
inputs.} 
\label{fig:r63}
\end{center}
\vspace*{-0.5cm}
\end{figure}

In the following, we consider the DRSR $r_{6/3}$ which we show in Fig.\,
\ref{fig:r63}. The optimal result is obtained for the set 
$(\tau,t_c)= (0.46,20)$  $(\rm GeV^{-2},\rm GeV^2)$ corresponding to 
the minimum of $\tau$ and $t_c$ for $r_{6/3}$. We obtain:
\begin{eqnarray}
    r_{6/3} &=& 0.9966(10) \nonumber\\
    M_{X_{c,6}} &=& 3863(76)~{\rm MeV},
\end{eqnarray}
where we have used the predicted mass in Eq. (\ref{eq:x3}).

\subsection*{Mass from the molecular currents using DRSR}
Finally, we deduced the masses of $D^*D$ and $J/\psi\pi$ molecular 
configuration, using DRSR. After the optimization procedure, the 
results are:
\begin{eqnarray}
    r_{\psi\pi/3} &=& 1.0034(7) \nonumber\\
    M_{X_{c,\psi\pi}} &=& 3889(76)~{\rm MeV},
    \\ && \nonumber\\
    r_{D^*D/3}&=&1 \nonumber\\
    M_{X_{c,D^*D}} &=& M_{Z_{c,D^*D}}=3912(61)~{\rm MeV},
\end{eqnarray}
where we have used the predicted mass in Eq. (\ref{eq:x3}).

\subsection*{${\cal T}_{X_c}$ tetramole}
Taking the fact that the different assignments to $X_c$ lead to almost 
degenerated states and almost the same coupling to the currents, we 
consider that the observed state is their combination which we call 
tetramole ${\cal T}_{X_c}$ with the mean mass and coupling\,:
\begin{eqnarray}
    M_{{\cal T}_{X_c}} &=& 3876(44)~{\rm MeV} \nonumber\\
    f_{{\cal T}_{X_c}} &=& 183(16)~{\rm keV}.
\end{eqnarray}

\section {The b-sector}
We extend our previous analysis for the $b$-quark states. Essentially, 
the analysis can be summarized exchanging the $c$-quark by the 
$b$-quark in the interpolating currents. The results obtained with the 
LSR at NLO, the DRSR and the optimization techniques allow us to 
establish the results summarized in Table \ref{tab:bsector}.

\begin{table}[hbt]
\setlength{\tabcolsep}{0.27pc}
{
\begin{tabular}{cccc}
\hline
\hline
  States & Masses (Mev) & Couplings (keV) & Data \\
\hline
  $Z_b$ & $10579(99)$ & $10(2)$ & $Z_b(10650)$\,\cite{BELLEZb} \\
  $X_b$ & $10545(131)$ & $14(3)$ & -- \\
  $T_{bb\bar{u}\bar{d}}$ & $10501(98)$ & $33(7)$ & -- \\
  $T_{bb\bar{u}\bar{s}}$ & $10521(83)$ & $21(7)$ & -- \\
\hline
\end{tabular}}
 \caption{{\small Summary of results of the $XZT$ states in the b-sector.}}
 \vspace*{0.25cm}
\label{tab:bsector}
\end{table}

\section{Conclusions}
From the previous discussions, one can notice that the sum rules 
reproduce quite well the experimental masses of the $X_c(3872)$ and 
$Z_c(3900)$ within the molecules or/and four-quark state configurations. 
The DRSR has improved the accuracy of the predictions compared to the 
previous ones in the literature\,\footnote{For reviews on previous LO 
QCD spectral sum rules results, see e.g.\,
\cite{MOLEREV,ZHUREV,RAPHAEL, STEELE}.}. 

However, as quoted in \cite{DRSR11a}, the isolated study of the mass of 
the $X_c$ and $Z_c$ cannot provide a sharp selection for the four-quark 
and/or molecule nature of these states without studying in details their 
decay modes. At the present stage, we can only provide a description of 
these states as {\it tetramole} ({$\cal T$}) states.

We confront our results from LSR $\oplus$ DRSR with the ones from 
different approaches in the literature (lattice calculations\,
\cite{JUN,MALT,MOHAN,LESK}, light front holographic\,\cite{DOSCH2}, quark 
and potential models $\oplus$heavy quark symmetry\,
\cite{MENG,ROSNER,QUIGG,BRAATEN,CHENG,RICHARD2,ZHUMODEL,WU,BICUDO}). 
We observe that, there is (almost) a consensus for the predictions of the 
axial-vector $1^+$ masses from these approaches: the $T_{cc}$ state is 
expected to be around the physical threshold while the $T_{bb}$ one is 
below the threshold and then stable against strong interactions. 
However, the recent LHCb data for the $1^+~T_{cc}$ candidate does not 
favour the models of\,\cite{LUCHA,BRAATEN,CHENG} which predict a too high 
$1^+~T_{cc}$ mass. 

In general, from our approach, the masses of all states studied in this 
work are grouped around the respective physical thresholds.


\end{document}